\documentclass[preprint,showpacs,preprintnumbers,amsmath,amssymb]{revtex4}
\usepackage{graphicx}
\usepackage{dcolumn}
\usepackage{bm}
\usepackage{color}

\newcommand{\be}{\begin{eqnarray}}
\newcommand{\ee}{\end{eqnarray}}
\newcommand{\bi}{\bibitem}

\newcommand{\benl}{\begin{eqnarray*}}
\newcommand{\eenl}{\end{eqnarray*}}

\begin{document}
\title{Stability of Biaxial Nematic Phase for Systems with
       Variable Molecular Shape Anisotropy }
\author{Lech Longa$^{1,2,}$}\email[e-mail address:]{longa@th.if.uj.edu.pl}
\author{Grzegorz Paj\c{a}k$^{1}$}
\author{Thomas Wydro$^{1,2,}$}
\email[e-mail address:]{wydro@univ-metz.fr}

\affiliation{$^{1}$Marian Smoluchowski Institute of Physics,
Department of Statistical Physics and  Mark Kac Center for Complex
Systems Research, Jagiellonian
University, Reymonta 4, Krak\'ow, Poland\\
$^{2}$Universit\'e Paul Verlaine -- Metz, Laboratoire de Physique
Mol\'{e}culaire et des Collisions, 1 bvd Arago, 57078 Metz, France}
\date{\today}
\begin{abstract}
We study the influence of  fluctuations in  molecular shape on
the stability of the biaxial nematic phase by generalizing the mean
field model of  Mulder and Ruijgrok [Physica A {\bf 113}, 145
(1982)]. We limit ourselves to the case when the molecular
shape anisotropy, represented by the alignment tensor, is
a random variable of an annealed type. A prototype of such behavior
can be found in lyotropic systems - a mixture of potassium laurate,
1-decanol, and $D_2O$, where distribution of  the micellar shape
adjusts to actual equilibrium conditions. Further examples of
materials with the biaxial nematic phase, where molecular shape is
subject to fluctuations, are thermotropic materials composed of
flexible trimeric- or tetrapod-like molecular units. Our
calculations show that the Gaussian equilibrium distribution of the
variables describing molecular shape (dispersion force)
anisotropy gives rise to new
classes of the phase diagrams, absent in the original model. Depending
on properties of the shape fluctuations, the stability of the biaxial
nematic phase can be either enhanced
or depressed, relative to the uniaxial nematic phases. In the former case the
splitting of the Landau point into two triple points with a direct phase transition line from isotropic to biaxial phase is observed.
\end{abstract}
 \maketitle
\section{Introduction}
The biaxial nematic phase is one of the perennially challenging
problems of experimental soft matter physics. Although predicted
theoretically by Freiser in 1970 \cite{Freiser} it was not until
1980 that the first experimental observation of this phase has been
reported \cite{YuSaupe80}.  The  phase was discovered  in a
lyotropic, ternary  mixture of potassium laurate, 1-decanol, and
$D_2O$ and its stability attributed to shape change of the micellar
aggregates as function of temperature and concentration of
amphiphilic molecules \cite{Neto2005}. The search for a thermotropic
biaxial nematic has proved highly controversial for more than two
decades. A  first qualitative report on a stable uniaxial-to-biaxial
nematic phase transition has been reported by Li \emph{et. al.}
\cite{Rosenblat} in their system of flexible, ring-shaped trimeric
liquid crystal. Recently, this phase has also been  detected in two
classes of thermotropic materials. It was reported in
'banana-shaped' mesogens \cite{rexpbia02,rexpbia03} and in liquid
crystalline tetrapodes \cite{rexpbia04}. The last class of materials
is particularly interesting for it comprises molecules with four
mesogenic units connected together through a flexible spacer. The
optimal packing of such tetrapodes in the biaxial nematic phase is
achieved for a quasiflat, platelet configuration of the tetrapod's
mesogens that are, on the average, tilted in the plane of the
platelet.

A challenge for theory is to find molecular factors responsible for
absolute stability of the observed biaxial nematic phase. The
presently existing microscopic models
\cite{Freiser,Mulder,Mulder0,Longa1,Sonnet03} show that the
molecular shape- and  pair interaction biaxiality are obviously
important  for the biaxial phase to exist. However, as numerous
experimental reports have demonstrated \cite{Bruce}, they seem not
sufficient to get the absolutely stable biaxial phase. In the
present paper we show that a variable (fluctuating) anisotropy of
the molecular shape can be an important stabilizing factor for this
phase to occur. Indeed, it seems that this factor appears commonly
in the micellar- trimeric- and tetrapod systems. Let us mention that
the theoretical studies and computer simulations so far have
addressed the size and shape of the micelles  in lyotropic systems
\cite{NelsonRutledge97,GirardiFigueiredo03,ChristopherOxtoby03,DuqueTarazona97},
but a connection between  the shape anisotropy distribution and the
stability of the biaxial nematic phase  have not  yet been explored.

 The present  paper is arranged as follows.
 After introduction of the model in Sec. II,
we explore stability of the biaxial nematic phase on shape fluctuations in Sec. III.  Section IV is  devoted  to a short summary.

\section{The model}

We assume that the hamiltonian $H(\{\mathbf{\mu}\},\{\mathbf{S}\})$
of $N$ liquid crystalline molecules with dynamical degrees of
freedom $\{\mathbf{S}\}$ also depends   on the $ \{\mathbf{\mu}\}$
variables, which parameterize molecular shape. The partition
function is calculated for each allowed configuration of randomly
chosen $\{\mathbf{\mu}\}$. Then, in analogy to statistics of
disordered systems with annealed disorder \cite{annealed01},
$\{\mathbf{\mu}\}$ is treated as a set of dynamical variables that
adjust to actual equilibrium conditions. Under these circumstances
the free energy, $F$, of the system can be  approximated  by the
logarithm of the $\{\mathbf{\mu}\}$ averaged partition function
\begin{eqnarray}\label{F}
    F&=&-k_BT\ln Z  \\
    Z&=& \sum_{\{\mathbf{\mu}\}}P(\{\mathbf{\mu}\})\sum_{\{\mathbf{S}\}}
    e^{-\beta H(\{\mathbf{\mu}\},\{\mathbf{S}\}) }\label{Z},
\end{eqnarray}
where $P(\{\mathbf{\mu}\})$ is the probability distribution of the
$\{\mathbf{\mu}\}$ variables.
A role played by the distribution $P$ on the formation of the
biaxial nematic phase is studied by generalizing a very elegant
mean-field model of Mulder and Ruijgrok \cite{Mulder} (MR), which employs a
connection between the $\mathcal{SU}(3)$ symmetry and the biaxial
nematic liquid. The most important feature of the model is that its
partition function can be calculated exactly,
which, as we are going to show, allows one for a
semi-analytical treatment of the annealed average (\ref{Z}). More
specifically, in the MR model the internal, dynamical state of each
molecule is parameterized by eight degrees of freedom: three
components $L_\alpha$ of the angular momentum $\mathbf{L}$ and five
components $Q_{\alpha\beta}$ of the symmetric and traceless
quadrupole moment $\mathbf{Q}$. These eight variables are next
combined to form eight generators $S_a$ of the $\mathcal{SU}(3)$
algebra, establishing equivalence between $\{ \mathbf{L},\mathbf{Q}
\}$ and $\mathbf{S} = \sum_{a=1}^8 S_a \boldsymbol{\lambda}_a$,
where $\boldsymbol{\lambda}_a$ are the Gell-Mann matrices:
$(\mathbf{S})_{\alpha \beta}=Q_{\alpha \beta} - \frac{i}{2}
\sum_{\gamma}\varepsilon_{\alpha \beta \gamma} L_{\gamma}$,
$\alpha,\beta,\gamma=1\ldots3$.  The MR Hamiltonian is
the mean-field (MF) version  of the  Heisenberg-type interaction
\cite{Mulder}
\begin{equation}
H\equiv H_{MF}= \frac{J N}{4}\mathrm{Tr}(\bar{\mathbf{S}}
\bar{\mathbf{S}}) -  \frac{J}{2} \sum_{i=1}^N
\mathrm{Tr}(\bar{\mathbf{S}} \mathbf{S}_{i}), \label{MRhamiltonian}
\end{equation}
where the dynamical variables $\mathbf{S}_{i}$ are subject to two
$i$-independent constrains, represented by Casimir invariants of the
$\mathcal{SU}(3)$ algebra
\begin{eqnarray}
  \mathrm{Tr} (\mathbf{S}_{i}^2)&=&
\mathrm{Tr} (\mathbf{\tilde{Q}}_{i}^2)= \sum_{\alpha=1}^3
\mu_{i,\alpha}^2=2 I_2(\boldsymbol{\mu}_i) \label{MRrestrictions2}\\
  \mathrm{Tr} (\mathbf{S}_{i}^3)&=&
\mathrm{Tr} (\mathbf{\tilde{Q}}_{i}^3)= \sum_{\alpha=1}^3
\mu_{i,\alpha}^3=2 I_3(\boldsymbol{\mu}_i) \label{MRrestrictions3}
\end{eqnarray}
with  $\mu_{i,1}+\mu_{i,2}+\mu_{i,3}=0$; $N$ is the number of
molecules and $\bar{\mathbf{S}}$ is the thermodynamic average of $
\mathbf{S}_{i}$. The $\mu_{i,\alpha}$ variables are the eigenvalues
of the traceless matrices $\mathbf{S}_{i}$, or
$\mathbf{\tilde{Q}}_{i}$, where $\mathbf{\tilde{Q}}_{i}$ is the the
quadrupole moment of the $i-th$ molecule at rest ($\mathbf{L}_i
{=} 0$), obtained from
$\mathbf{S}_{i}$ by applying an $\mathcal{SU}(3)$ transformation.
Hence, if $\mathbf{\tilde{Q}}_{i}$ represents \emph{e.g.} the quadrupole moment
of a mass distribution the intrinsic properties of a given molecule,
like the ratios of its principal axes, enter through the Casimir
invariants $I_2(\boldsymbol{\mu}_i)$, $I_3(\boldsymbol{\mu}_i)$,
Eqs.~(\ref{MRrestrictions2},\ref{MRrestrictions3}). This remains in
full analogy with what we practise for an ordinary quadrupolar
tensor, where the   $I_2$- and $I_3$-like invariants are used to
characterize biaxiality of relevant physical observable
\cite{Gramsberger,longa3}. More specifically, depending on the
values of $I_2$ and  $I_3$, or their ratio
\begin{equation}
w=\sqrt{3}\frac{I_{3}}{(I_{2} )^{\frac{3}{2}}} \label{wm}
\end{equation}
three possibilities can be distinguished: (a) for $I_2= I_3=0$ the
tensor is isotropic; (b) for ${3} I_3^2= I_2^{3} $ the tensor is
uniaxial and (c) for $3 I_3^2< I_2^{3} $ the tensor is biaxial with
maximal biaxiality being obtained for $I_3=0$. The sign of $I_3$
decides about whether the tensor is prolate (plus sign) or oblate
(minus sign). Respectively, $w=1$ ($w=-1$) and $|w|<1$ refers to
rod-like (disk-like) and biaxial molecules. By construction the MR model is $\mathcal{SU}(3)$ invariant with degrees of freedom  running over the group
manifold and  its free energy is given in an  analytical form as
derived by Itzykson and Zuber \cite{ItzZub}.

\begin{figure}[ht]
\begin{picture}(500,300)
  \put(10,-10){
   \includegraphics{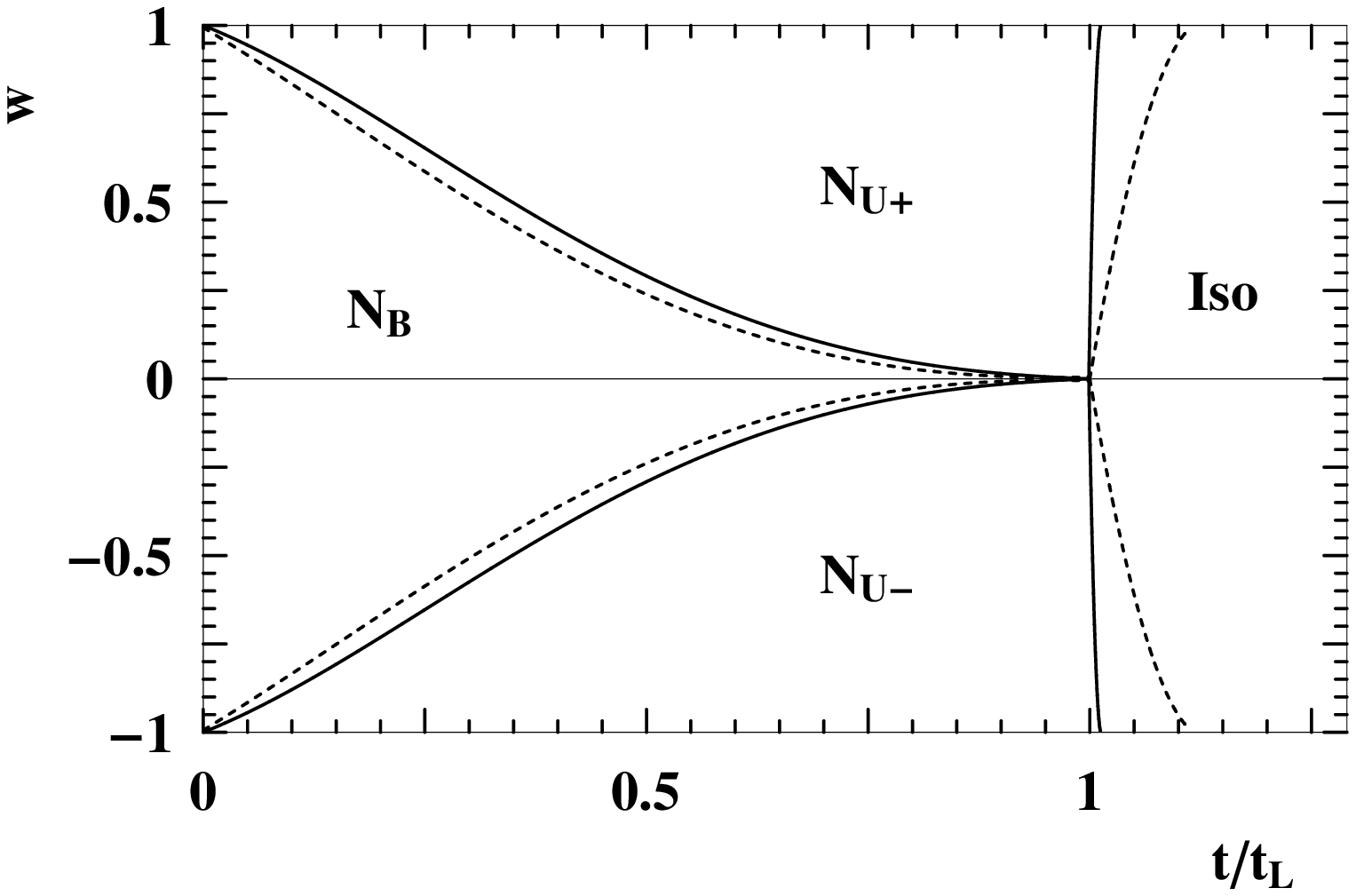}
  }
\end{picture}
\caption[]{ Comparison of the  MR  phase diagram \cite{Mulder}
(continuous lines) with one calculated  for the Luckhurst {\emph{et al.}} dispersion model \cite{rb05,Zannoni} (dashed lines); $-1 \le w=\sqrt{6}{\mathrm{Tr}(\mathbf{\hat{R}}^3)}
= \frac{1-6\kappa^{2}}{(1+2\kappa^{2})^{\frac{3}{2}}}\le 1$.} \label{Diagrams0}
\end{figure}

From physical point of view  the model matches the standard
Landau-deGennes phase diagram for biaxial nematics
\cite{Gramsberger},  known to characterize systems with purely
dispersion-type of interactions. It also reproduces the mean-field
results for the dispersion model of Luckhurst \emph{et al.}
\cite{rb05,Zannoni}, Fig.~\ref{Diagrams0}, given we take the pair
interactions of the form $V= -\epsilon\,
\mathrm{Tr}(\mathbf{\hat{R}} \,\mathbf{\hat{R}}')$, where
${\mathbf{\hat{R}}}$  denotes  the normalized quadrupole tensor
$(\mathrm{Tr}\,\mathbf{\hat{R}}^2=1)$ defined through the relation:
$\sqrt{1+2\kappa^{2}}\mathbf{\hat{R}}=\frac{1}{\sqrt{6}}(3\hat{\mathbf{z}}
  \otimes\hat{\mathbf{z}}-\mathbf{1})\pm \kappa (\hat{\mathbf{x}}
 \otimes\hat{\mathbf{x}}-\hat{\mathbf{y}}\otimes\hat{\mathbf{y}}) $;   $\kappa$ is the ratio of biaxial-to-uniaxial polarizability of the molecule. The relative error for the $N_U-N_B$ boundary calculated for both models does not exceed  2\% and is even smaller for the nematic order parameter (\emph{see} Fig.~4 in \cite{Mulder}).
Note however the usefulness of the $\mathcal{SU}(3)$ symmetry. It avoids calculation of integrals over Euler angles inherent to the dispersion model, which, in turn, allows for comprehensive studies of flexible quadrupoles.

   A generalization of the  MR model to systems with  variable anisotropy of
the molecular shape (dispersion forces) is straightforward. We  assume that $\mu_{i,1}$
and $\mu_{i,2}$  are annealed  degrees of freedom weighted  with the
probability $P(\mu_{1,1}, \mu_{1,2}, \mu_{2,1},...,\mu_{N,2})
\approx \prod_{i=1}^N P(\mu_{i,1}, \mu_{i,2})$. The mean-field
partition function, $Z=Z_1^N$, and the dimensionless free energy are
then given by
\begin{eqnarray}
Z_1 &=& e^{ -\frac{\mathrm{Tr}(\bar{\mathbf{S}}
\bar{\mathbf{S}})}{4t} }
\, Q \label{Z1}\\
Q &=& \langle \int d\mathbf{S}\,\,e^{
\frac{\mathrm{Tr}(\bar{\mathbf{S}} \mathbf{S})}{2t}}
\delta(\scriptstyle{\frac{\mathrm{Tr} (\mathbf{S}^2)}{2}}
\displaystyle - I_{2}(\boldsymbol{\mu})) \delta(\scriptstyle
\frac{\mathrm{Tr} (\mathbf{S}^3)}{2}\displaystyle-
I_{3}(\boldsymbol{\mu}))\rangle \,\,\,\,\,\,\,\,\,\,
\label{QQ}\\
F &=& -N t\, \log Z_1,\,\,\,\,\,\,\, \label{F-MF}
\end{eqnarray}
where $ t = k_B T/J$ is the dimensionless temperature and where
$\langle\,...\,\rangle = \int (...)P(\mu_{1},
\mu_{2})d\mu_{1}d\mu_{2}$. According to Itzykson and Zuber
\cite{ItzZub} the integral over $\mathbf{S}$ in (\ref{QQ}) can be
carried out to give
\begin{equation}
Q=\frac{2}{\Delta_{\boldsymbol{\bar{\gamma}}}}\left\langle
\frac{D}{\Delta_{\boldsymbol{\mu}}}\right\rangle, \label{Q1}
\end{equation}
where
\begin{equation}
D=\det\left(
\begin{array}{ccc}
e^{\bar{\gamma}_{1}\mu_{1}} & e^{\bar{\gamma}_{2}\mu_{1}} &
e^{\bar{\gamma}_{3}\mu_{1}} \\
  e^{\bar{\gamma}_{1}\mu_{2}} & e^{\bar{\gamma}_{2}\mu_{2}} &
  e^{\bar{\gamma}_{3}\mu_{2}} \\
  e^{\bar{\gamma}_{1}\mu_{3}} & e^{\bar{\gamma}_{2}\mu_{3}} &
  e^{\bar{\gamma}_{3}\mu_{3}}
\end{array}
\right),
\end{equation}
and where
\begin{equation}
\Delta_{\mathbf{x}}=(x_{1}-x_{2})(x_{2}-x_{3})(x_{3}-x_{1}).
\end{equation}
The eigenvalues $\bar{\gamma}_{\alpha}$ of the traceless matrix
$\bar{\mathbf{S}}/(2t)$, are determined from the minimum of the free
energy (\ref{F-MF}). In analogy to (\ref{MRrestrictions2},\ref{MRrestrictions3})
the invariants $ I_2(\boldsymbol{\bar{\gamma}})$ and
$I_3(\boldsymbol{\bar{\gamma}})$ are used to distinguish between (a)
the isotropic phase
($I_2(\boldsymbol{\bar{\gamma}})=I_3(\boldsymbol{\bar{\gamma}})=0$);
(b) the uniaxial nematic phase  ($3
I_3(\boldsymbol{\bar{\gamma}})^2=I_2(\boldsymbol{\bar{\gamma}})^3$)
and (c) the biaxial nematic phase ($3
I_3(\boldsymbol{\bar{\gamma}})^2<I_2(\boldsymbol{\bar{\gamma}})^3$).
In addition,  $I_3(\boldsymbol{\bar{\gamma}})$ is positive for
prolate uniaxial phase  ($N_{U+})$ and negative for oblate uniaxial
phase ($N_{U-})$.

The annealed averaging $\langle\,...\,\rangle$ over $P(\mu_{1},
\mu_2)$, Eq. (\ref{QQ}), has a very simple interpretation in the
mean-field theory. Setting $\bar{\gamma}_{\alpha}=0,\,\,
(\alpha=1,2)$, which is always one of the mean-field solutions, we
find that $P(\mu_{1}, \mu_2)$ matches the density distribution of
the molecular shape anisotropy in the reference (stable or
metastable) disordered phase. We believe therefore that for a
credible choice of $P(\mu_{1}, \mu_2)$ the model correctly
reproduces generic phase behavior for flexible quadrupoles in the
vicinity of the isotropic phase. Clearly, the original MR model is
recovered if $P(\mu_{1}, \mu_2)$ is given by Dirac delta
distribution. In what follows we take  $P$ to be the Gaussian
distribution. This choice is consistent with the maximum entropy
principle applied in the isotropic phase and the observation that
usually  only first two moments of $P$ can be determined reasonably
well from experiment \cite{Neto2005}. Assuming that in the reference
(disordered) phase these moments are given by:
$<\mu_\alpha>=m_\alpha$, $<(\mu_\alpha-m_\alpha)^2>=\sigma_\alpha^2$
and $<(\mu_1-m_1)(\mu_2-m_2)>=\tilde{\lambda}\sigma_1\sigma_2$
($\alpha=1,2$) we find
\begin{equation}\label{pdf}
    P(\mu_{1},\mu_2)= \frac{\sqrt{a}}{2\pi\sigma_1\sigma_2}
    e^{-\frac{1}{2}\sum_{\alpha\beta}(\mu_\alpha-m_\alpha)\sigma_{\alpha\beta}
    (\mu_\beta-m_\beta)},
\end{equation}
where
\begin{equation}\label{Gauss}
\boldsymbol{\sigma}=\left(
\begin{array}{cc}
\frac{a}{\sigma_1^{2}} &  -\frac{\lambda}{\sigma_1\sigma_2}\\
 -\frac{\lambda}{\sigma_1\sigma_2}  &  \frac{a}{\sigma_2^{2}}
\end{array}
\right)
\end{equation}
with  $\lambda$ being the real parameter,
$a{=}\frac{1}{2}(1+\sqrt{1+4\lambda^2})$
and $-1\le
\tilde{\lambda}{=}\frac{\lambda}{a}
\le 1$. The distribution $P(\mu_3)$ of $\mu_3$, obeying the
constrain $\mu_1+\mu_2+\mu_3=0$, is also a Gaussian with average
$<\mu_3>=m_3=-m_1-m_2$ and dispersion ${\sigma}_3^2=\sigma_1^2
+2\tilde{\lambda}\,\sigma_1\sigma_2 +\sigma_2^2 $ ($|\sigma_1 -
\sigma_2|  \le \sigma_3 \le \sigma_1 + \sigma_2$): $P(\mu_3)=
\frac{1}{\sqrt{2\pi}{\sigma}_3}
\exp(-{\frac{(\mu_3+m_1+m_2)^2}{2{\sigma}_3^2}})$. In general, the parameters of the distribution (\ref{pdf}) can depend on temperature, but this possibility, which can be  relevant for quantitative understanding of phase diagrams in lyotropic systems, will not be discussed here.

\section{Results}

The phase diagrams, obtained from the global minimization of the
free energy (\ref{F-MF}) with respect to $\{\bar{\gamma}_{\alpha}, \, \alpha=1,2\}$, depend on
the values of five parameters: ${m}_1,\,m_2$, ${\sigma}_1,\,\sigma_2$ and $\lambda$ (or $\sigma_3$)
of the Gaussian distribution (\ref{pdf}). To make a
direct comparison with the earlier work \cite{Mulder}  we use,
instead of ${m}_1$ and $m_2$, the molecular shape parameter
$w(\boldsymbol{m})$ and $I_2(\boldsymbol{m})$,
Eqs.~(\ref{MRrestrictions2},\ref{wm}). A connection between the
parameterizations is given by
\begin{eqnarray}\label{m1m2}
  m_1+m_2 &=& u  \\
  m_1 m_2 &=& u^2-I_2,\label{m1m22}
\end{eqnarray}
where  u is a solution of the cubic equation
\begin{equation}\label{cubic}
    u^3 - u I_2 + \frac{2\sqrt{3}}{9}(I_2)^{3/2}w =0.
\end{equation}
 The three real roots $u_1\le u_2 \le
u_3$ of Eq.~(\ref{cubic}) correspond to different permutations
between axes of the molecule-fixed frame. In what follows we choose
the $u_2$-solution to identify the corresponding $m_1$ and $m_2$.
Solving Eqs.~(\ref{m1m2},\ref{m1m22}) for given  $u_2$ and $I_2$
still leaves a freedom  to choose the pair $\{m_1,m_2\}$ up to a
permutation. We select the solution for which $m_1 \le m_2$.
Permutation symmetry of $Q$:
$Q(\{m_\alpha\},\{\sigma_\alpha\},...)=Q(\{m_{{\cal{P}}(\alpha)}\},
\{\sigma_{\cal{P}(\alpha)}\} ,...)$, where $\cal{P}$ is an arbitrary
permutation of $\{1,2,3\}$, allows us to construct the remaining
diagrams for $m_1 \ge m_2$ from the ones given.

Numerical calculations are carried out for fixed values of
${\sigma}_1,\,\sigma_2$ and $\lambda$. The phase diagrams are shown
in plane of the molecular shape parameter $w(\boldsymbol{m})$ and
the reduced temperature $t/t_L$, with $t_L \ge
\frac{<I_2(\boldsymbol{\mu})>}{8}$  being the isotropic-nematic
transition temperature for $w(\boldsymbol{m})=0$. In all cases the
numerical value of $I_2$ was fixed to $I_2= \frac{1}{4}$. The
diagram for  $I_2= \frac{1}{4} \xi^2$, $\sigma_i$ and $\lambda$,
where $\xi$ is an arbitrary real number, can be obtained from that
for $I_2= \frac{1}{4}$, $\sigma_i \rightarrow \sigma_i/\xi$ and
$\lambda$, which follows from  invariance of $Q$, Eq.~(\ref{QQ}),
with respect to  $\xi$-rescaling of the parameters: $Q( \{ m_i\},
\{\sigma_i\}, \lambda, t )$ $=$ $Q(\{\xi m_i\}, \{ \xi \sigma_i\},
\lambda, \xi t )$. In addition, the invariance of $Q$, Eq.(\ref{QQ}), with
respect to change of $\{\boldsymbol{\mu}, {\boldsymbol{\bar{\gamma}}}\}$
into  $\{-\boldsymbol{\mu}, -{\boldsymbol{\bar{\gamma}}}\}$ makes the
phase diagrams symmetric with respect to the line  $w(\boldsymbol{m})=0$.

\begin{figure}[ht]
\begin{picture}(500,300)
  \put(10,-10){
   \includegraphics{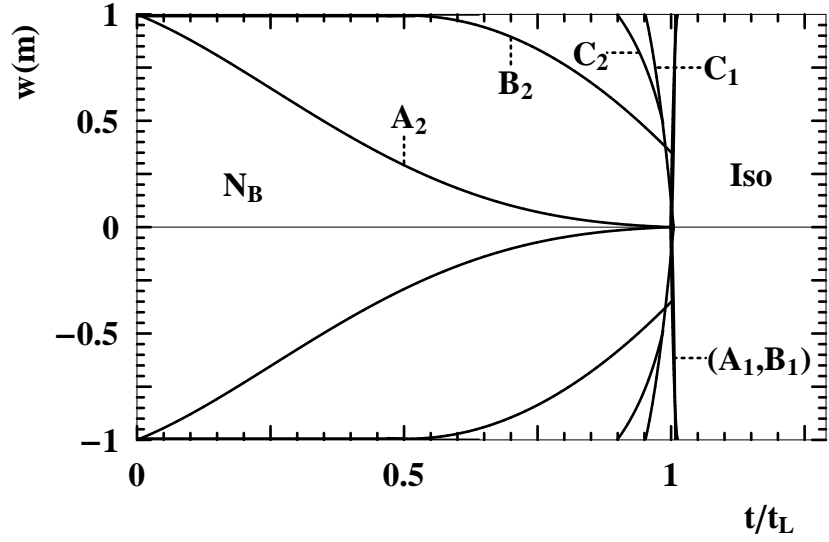}
  }
\end{picture}
\caption[]{ Phase diagrams  for $\sigma_{2}=0$. Diagrams labeled
$A_i$, $B_i$ and $C_i$ correspond  to $\sigma_{1}=$ 0 \cite{Mulder},
0.15 and 0.3, respectively. Subscript $i=1$ refers to the isotropic
(Iso)-nematics phase transition lines whereas subscript $i=2$ refers
to the  uniaxial nematic-biaxial nematic ($N_B$) lines.}
\label{Diagrams1}
\end{figure}

\begin{figure}[ht]
\begin{picture}(500,300)
  \put(10,-10){
   \includegraphics{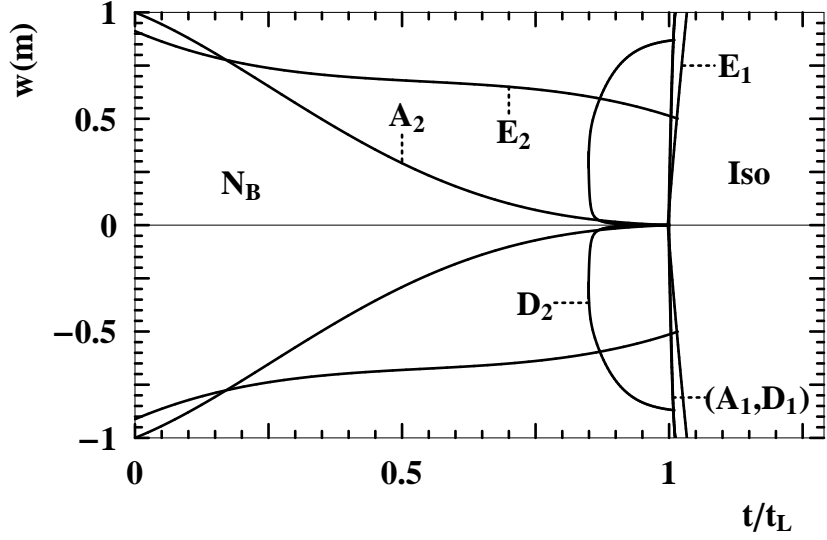}
  }
\end{picture}
\caption[]{ Phase diagrams  for $\sigma_{1}=\sigma_{2}$. Diagram
labeled $A_i$ corresponds to $\sigma_{1}=$ 0 \cite{Mulder}, while
diagrams $D_i$ and $E_i$   to $\sigma_1=0.15$.  Subscript $i=1$
refers to the isotropic (Iso)-nematics phase transition lines
whereas subscript $i=2$ refers to the uniaxial nematic-biaxial
nematic ($N_B$) lines. For diagrams $D_i$ and $E_i$ we took
$\tilde\lambda= 0.99$ and $\tilde\lambda= -0.99$, respectively.}
\label{Diagrams2}
\end{figure}

Numerical minimization of the free energy allows us to divide all the diagrams
into classes shown in Figs.~\ref{Diagrams1}-\ref{Diagrams3}.  The
corresponding isotropic-nematic transition temperature $t_L $ is
plotted in Figs.~\ref{tLandau1},\ref{tLandau2}. At high
temperatures, in the vicinity of isotropic-nematic phase transition,
the results can be understood qualitatively   from the expansion of
$Q$, Eq.~(\ref{Q1}), about $\bar{\gamma}_i =0$. Up to sixth-order in
$\bar{\gamma}_i$ it reads
\begin{eqnarray}\label{Q-expansion}
    Q &\approx& 1+ \frac{1}{4}\, \langle \,I_2\,\rangle\,
          I_2(\bar{\boldsymbol{\gamma}}) +
\frac{1}{10} \,\langle\, I_3\,\rangle\,
          I_3(\bar{\boldsymbol{\gamma}}) +
\frac{1}{40} \,\langle \,I_2^2\,\rangle\,
          I_2(\bar{\boldsymbol{\gamma}})^2 \nonumber \\
          &+&
          \frac{1}{70}\, \langle\, I_2\, I_3\,\rangle\,
          I_2(\bar{\boldsymbol{\gamma}})\,I_3(\bar{\boldsymbol{\gamma}})
          +
\left(\frac{17}{12096}\, \langle \, I_2^3\,\rangle -
 \frac{1}{5040}\, \langle I_3^2\,\rangle\right)
          I_2(\bar{\boldsymbol{\gamma}})^3\nonumber \\
          &+&
\left(\frac{1}{420}\,\langle \, I_3^2 \,\rangle -
 \frac{1}{5040}\,\langle\, I_2^3\,\rangle\right)
          I_3(\bar{\boldsymbol{\gamma}})^2 + \ldots,
\end{eqnarray}
where the averages over  $\boldsymbol{\mu}$ are given by
\begin{eqnarray}\label{averages-I}
  \langle\, I_2\,\rangle   &=& I_{20} + I_{02} \nonumber\\
  \langle\, I_3\,\rangle   &=& I_{30} + 3 I_{12}\nonumber\\
  \langle\, I_2^2\,\rangle &=&I_{20}^2 + 6 I_{22} + 3 I_{04}\nonumber\\
\langle\, I_2 \,I_3\,\rangle &=& I_{20} I_{30} +6 I_{20}  I_{12} +
                           4 I_{30} I_{02} + 9 I_{14} \\
\langle\, I_3^2\,\rangle   &=& I_{30}^2 + 18 I_{02}^3 + \frac{45}{2}
 I_{06} + 15 I_{12} (I_{30} + 3 I_{12} )\nonumber \\ &&
 + \frac{9}{2} I_{02} ( I_{20}^2
 + 2 I_{22} + 2 I_{02} I_{20} - 9 I_{04}) - 18 I_{04} I_{20}
 - \frac{9}{2} I_{20} I_{22} \nonumber \\
 \langle\, I_2^3\,\rangle &=&I_{20}^3 + 3 I_{02} ( 9 I_{04}
   -4 I_{02}^2   - 12 I_{02} I_{20} - I_{20}^2 )+ 27 I_{04} I_{20}
   + 18 I_{22} (3 I_{02} + I_{20}), \nonumber
\end{eqnarray}
with
\begin{equation}\label{averages-def}
    I_{pq}= \frac{1}{2}\sum_{\alpha=1}^3 m_\alpha^p\, \sigma_\alpha^q
    \hspace{2cm} (I_2 \equiv I_{20}, \, I_3 \equiv I_{30}).
\end{equation}

 Using Eqs.~(\ref{F-MF},\ref{Q-expansion})
we find that in the limit of small $\{\bar{{\gamma}}_i\}$ the
bifurcation from isotropic to nematic phase takes place at
$t_b{=}\frac{1}{8} \langle\,
I_2\,\rangle$. Due to the $I_{02}$ contribution   to $\langle\,
I_2\,\rangle$ $(I_{02}\ge 0)$ the bifurcation temperature, $t_b$, is
always greater than the corresponding bifurcation temperature for
the mono-dispersive system with molecules characterized by the
average shape parameters $\{m_i\}$. Similar result should hold for
the transition temperatures, for they usually follow behavior of
$t_b$.

In the vicinity of the isotropic phase  the terms higher than sixth
order  in $\{\bar{{\gamma}}_i\}$ can be neglected in the expansion
(\ref{Q-expansion}). Out of the six terms that are left one can
associate $ \langle\, I_3\,\rangle$ with an effective molecular
shape anisotropy of the system in the isotropic phase. The $I_{12}$
term, contributing to $ \langle\, I_3\,\rangle$ and being  of
undetermined sign, effectively changes this anisotropy and thus has
a profound effect on stability of the biaxial phase. More
specifically, as $ \langle\, I_2\,\rangle \ge I_{20}$ and $
\langle\, I_2^2\,\rangle \ge I_{20}^2$, for $I_{30}$ and  $I_{12}$
of opposite sign with $|I_{12}| < |I_{30}|$ the
$\langle\,I_3\,\rangle$ coefficient can effectively be reduced by
the shape fluctuations. If, simultaneously, the invariant $\langle\,
I_2 I_3\,\rangle$ is also reduced  by fluctuations, which as we
checked is easily achieved in the parameter space, the range of
stability of the biaxial phase in ($w(\mathbf{m}), t/t_L$) space
gets enhanced as compared to the case without shape fluctuations.
That is, a sufficient condition to observe a constructive role of
shape fluctuations in stabilizing the biaxial nematic phase is the
simultaneous fulfilment of two inequalities
\begin{eqnarray}\label{inequalities}
|\langle\, I_3\,\rangle| &\le& |I_{30}| \\
|\langle\, I_2 I_3\,\rangle| &\le& I_{20} |I_{30}|.
\end{eqnarray}
Interestingly, any shape fluctuations about spherically
symmetric shape ($m_\alpha =0$) stabilize  biaxial nematic phase of
maximal biaxiality ($\langle\, I_2^n I_3\,\rangle=0$), without
intermediate uniaxial phase.

Now we turn to detailed analysis of the model. We carried out numerical minimization to determine phase diagrams for a representative set of model parameters.  All distinct classes of the diagrams
identified are shown in Figs.~\ref{Diagrams1}-\ref{Diagrams3}. In particular, we found
that the class of parameters where the biaxial nematic phase enhances its stability is much more reacher that the condition (\ref{inequalities}) may suggest. However, we are unable to find numerical limitations on the model parameters  in a compact form, except for  some limiting cases. But even in these limiting cases we recover all observed classes of the phase diagrams.   The simplest case occurs when one of the dispersions, say $\sigma_2$, vanishes. This
corresponds to the Gaussian distribution for $\mu_1$ and $\mu_3$
($\sigma_1=\sigma_3$), and Dirac delta distribution for $\mu_2$.
Phase diagrams, influenced by this polydispersivity, already exhibit quite
different topology as compared to the original MR model. This we
illustrated in
Fig.~\ref{Diagrams1},
where diagrams B and C correspond to $\sigma_{1}=0.15$ and
$\sigma_{1}=0.3$, respectively. Note a considerable enhancement of
stability of the biaxial phase along with splitting of  the original
MR quadruple Landau point into two triple points. The triple points
are connected by the line of the first order transition between the
isotropic and biaxial phases. Moreover, as expected, the transition
temperature, $t_{L}$, of the isotropic-nematic phase transition
increases with increasing value of $\sigma_{1}$,
Fig.~\ref{tLandau1}.

Subsequent case to consider is the full Gaussian distribution. We
limit ourselves to the symmetric distributions with
$\sigma_{1}=\sigma_{2}$. Now the phase diagrams exhibit yet another
topology, which is represented by the diagram D in
Fig.~\ref{Diagrams2}. Amazingly, there are two triple points and
the Landau point on one diagram, and two lines of the direct first
order phase transition between the isotropic and biaxial phases for
$|w(\mathbf{m})|\gtrapprox 0.8$.
 Stable uniaxial phases form two bubble-like islands.
As before, the transition temperature, $t_{L}$, between the
isotropic- and nematic phases is higher than that of the MR model
and increases with increasing $\tilde{\lambda}$,
Fig.~\ref{tLandau2}. Another possible diagrams obtained for this
case are shown in Fig.~\ref{Diagrams3}. In the diagram F the biaxial
phase is practically eliminated being reduced to a line
$w(\mathbf{m})=0$. It is interesting to follow reduction of
stability of the biaxial phase by  comparing diagrams B, H, G and F,
where changes of $\sigma_{2}$ from zero to $\sigma_{1}$, for fixed
$\sigma_{1}$, correspond to successive phase diagrams. Numerical
estimates of low-temperature part of the phase diagrams have been
checked  to stay consistent with asymptotic expansion
($\frac{1}{t}\rightarrow\infty$) of the selfconsistent equations
$\{\partial F / \partial \bar{{\gamma}}_i=0 \}$ for
$\{\bar{{\gamma}}_i\}$.

\section{Summary}

We have studied the influence of the variable molecular
shape anisotropy  on stability of the biaxial nematic phase. To make
the analysis as simple as possible we generalized the exact mean
field solution obtained by Mulder and Ruijgrok \cite{Mulder} for
biaxial molecules to the case when the quadrupole tensor is a
dynamical variable. We assumed that at equilibrium, the molecular
shape anisotropy can be approximated by the (annealed) distribution,
$P$, of the molecular parameters $\{\mu_\alpha \}$. In the
mean-field approximation the natural choice for $P$, consistent with
the maximum entropy principle applied in the isotropic phase, is the
two-dimensional Gaussian distribution.

The nonzero second moments of the Gaussian distribution lead to
important remodeling of the original MR phase diagram. We observe
that polydispersivity changes the range of stable biaxial phase and
that behavior of the system can qualitatively differ from its
mono-dispersive counterpart characterized by the average shape
parameters $\{m_i\}$.  Generally, the transition between the
isotropic and the nematic phases occurs at higher temperatures when
molecular shape changes are allowed. The phase diagram is modified,
for instance, by showing the quadruple Landau point being splitted
into two triple points connected by a line of first order
transitions between the isotropic and biaxial phases. By comparing
diagrams B, D and E we can conclude that strong correlations between
shape fluctuations along main molecular axes
($|\tilde{\lambda}|\rightarrow 1$), or fluctuations along two of the
three molecular principal axes lead to particularly large region of
stable biaxial phase. Importantly, fluctuations about isotropic
shape give rise to stable biaxial nematic without intermediate
uniaxial phase, while fluctuations about fixed $w(m)$, diagram G,
show on the temperature axis two uniaxial phases separated by the
biaxial nematic.

\begin{figure}[ht]
\begin{picture}(500,300)
  \put(10,-10){
   \includegraphics{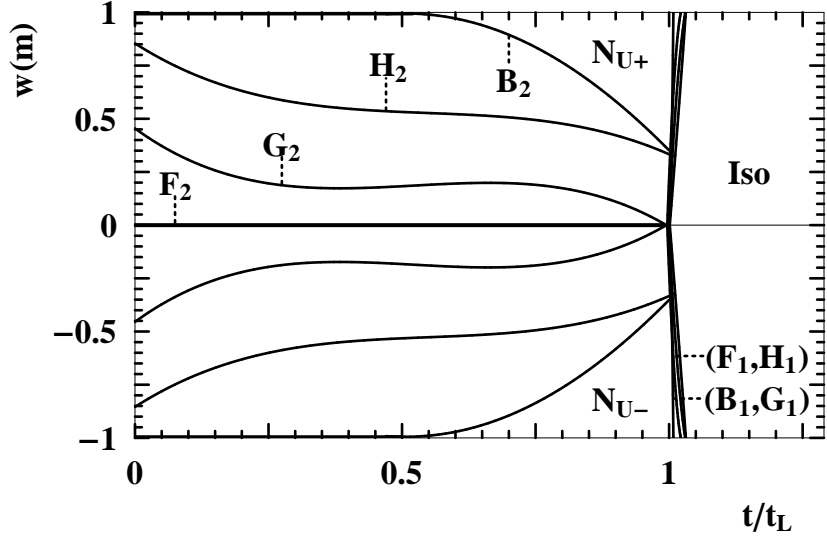}
  }
\end{picture}
\caption[]{  Phase diagrams for $\sigma_{1}=0.15$ and $\lambda=0$.
Diagram $B_i$ corresponds to $\sigma_{2}=0$, while diagrams $F_i$,
$G_i$ and $H_i$ correspond to $\sigma_{2}=$ 0.15, 0.1 and 0.05,
respectively. Subscript $i=1$ refers to the isotropic (Iso)-nematics
phase transition lines whereas subscript $i=2$ refers to the
uniaxial nematic ($N_U+, N_U-$)-biaxial nematic lines. The lines of
isotropic-nematic phase transitions occur in the following order
(from left to right): $B_{1}$, $G_{1}$, $F_{1}$ and $H_{1}$.}
\label{Diagrams3}
\end{figure}

\begin{figure}[ht]
\begin{picture}(500,300)
  \put(10,-10){
   \includegraphics{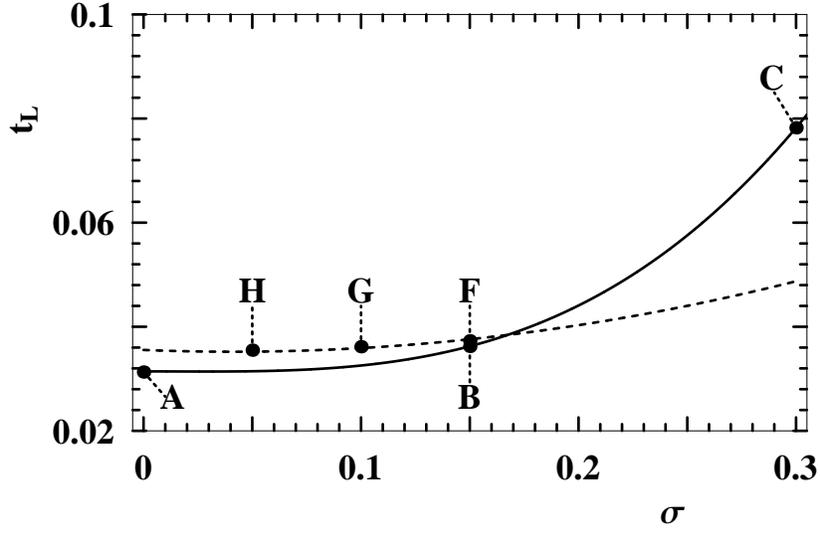}
  }
\end{picture}
\caption[]{Isotropic-nematic  transition temperature $t_{L}$ for
$w(\mathbf{m})=0$ as  function of $\sigma_1$ or $\sigma_2$,
collectively denoted  $\sigma$. For continuous line $\sigma
=\sigma_{1}$ with $\sigma_{2}=0$ whereas for dashed line $\sigma
=\sigma_{2}$, $\sigma_{1}=0.15$ and $\lambda=0$. Points A, B and C
correspond to $\sigma_{1}=$ 0, 0.15 and 0.3, respectively. Points F,
G and H correspond to $\sigma_{2}=$ 0.15, 0.1 and 0.05,
respectively.} \label{tLandau1}
\end{figure}

\begin{figure}[ht]
\begin{picture}(500,300)
  \put(10,-10){
   \includegraphics{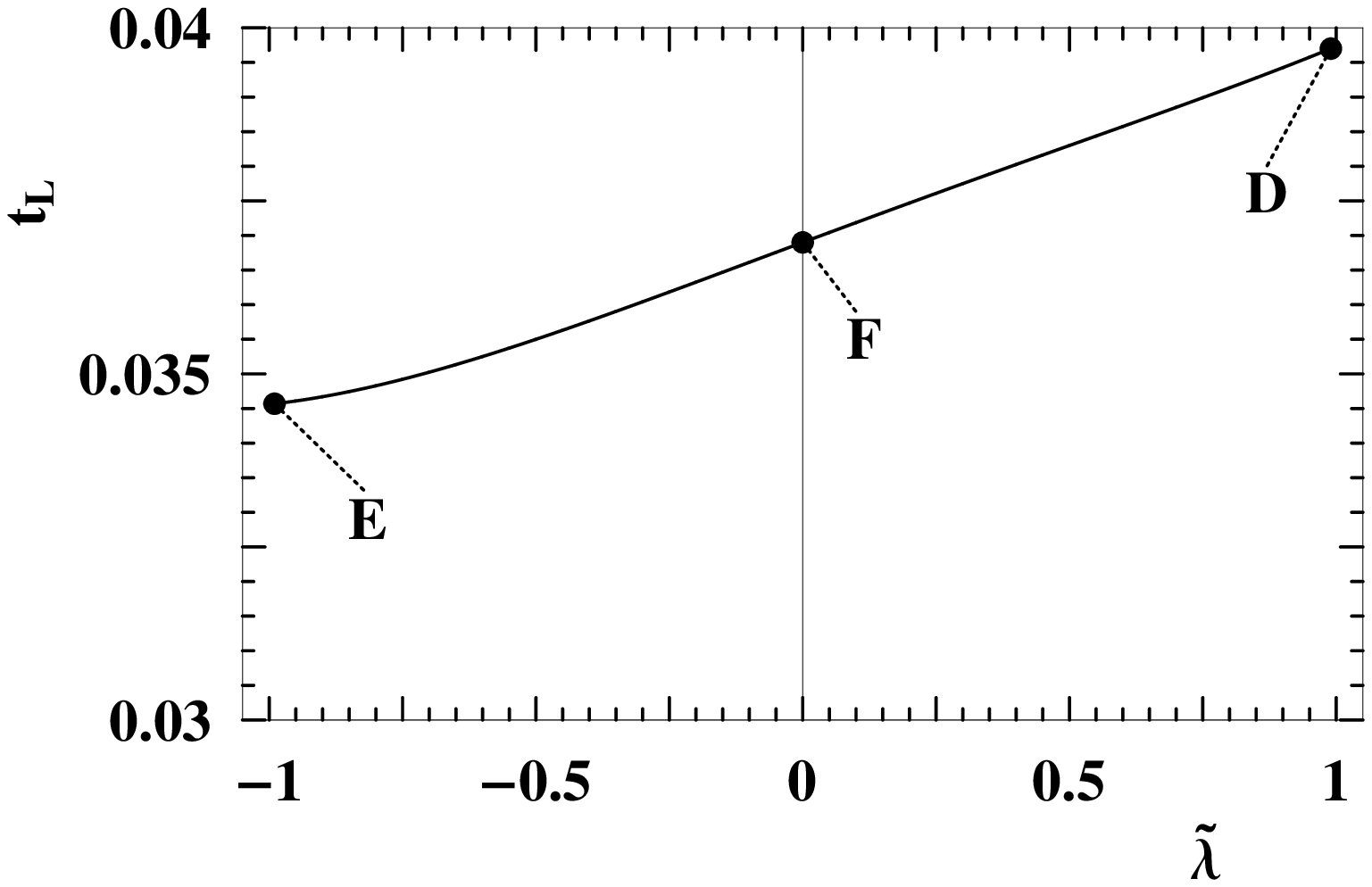}
  }
\end{picture}
\caption[]{Isotropic-nematic  transition temperature $t_{L}$ for
$w(\mathbf{m})=0$ as  function of $\tilde\lambda$ for
$\sigma_{1}=\sigma_{2}$. Points D, E and F correspond to
$\tilde\lambda=$ 0.99, -0.99 and 0, respectively.} \label{tLandau2}
\end{figure}

In some  cases the biaxial phase can be destabilized in the vicinity
of the isotropic phase giving only the uniaxial nematic phases and
first-order phase transitions between them (class F,
Fig.~\ref{Diagrams3}). Similar case has recently been observed by
Bates \cite{Bates} in his computer simulation of a generic, flexible
V-shaped molecules on a lattice. The only difference between our
predictions and that of \cite{Bates} is that we do not observe a
biaxial nematic phase at low temperatures, shown in Fig.~5(a) of
\cite{Bates}. A reason for that is our neglecting of temperature
dependence  in the isotropic distribution (\ref{pdf}) at low
temperatures \cite{explanation}. Clearly, to be consistent with
general thermodynamics at $T=0$ the distribution (\ref{pdf}) should
approach Dirac delta function about $m_\alpha$. In our studies we
have disregarded any temperature dependence in (\ref{pdf}), being
primarily interested in system's behavior close to the isotropic
phase. However, the diagram predicted in  \cite{Bates} can also be
obtained within our model if we  replace $\sigma_\alpha$ by
$\sqrt{t} \sigma_\alpha$ in (\ref{pdf}). Then the ground state of
(\ref{F-MF}) for $w(\mathbf{m})\ne \pm 1$ would always be the
biaxial nematic phase and, consequently, the phase diagrams of the
class F, Fig.~\ref{Diagrams3}, would go into generic diagrams found
by Bates.

\begin{acknowledgments}
This work was supported by Grant N202 169 31/3455 of the Polish
Ministry of Science and Higher Education, and by the EC Marie Curie
Actions 'Transfer of Knowledge', project COCOS (contract
MTKD-CT-2004-517186).
\end{acknowledgments}

\newpage
%

\newpage
%

\newpage
%

\newpage
%

\newpage
%

\newpage

%


\begin{thebibliography}{999}
\bibitem{explanation} By this we mean that
the mean-field theory of the biaxial nematics
renders the differences between the fluctuating $C_{2v}$-symmetric model
of Bates and the fluctuating $D_{2h}$-symmetric  quadrupole model
of secondary importance. Indeed, take  the $D_{2h}$-symmetric one-particle distribution
function as given by Eq.~35 in \cite{Longa1}. Then, the mean-field approximation
to the model of Bates leads to
an effective $D_{2h}$-symmetric, mean-field
interaction consistent with one under curly brackets
of Eq.~(37) in \cite{Longa1}. Shape fluctuations
go into fluctuations in the parameters of the effective potential.
Hence, the 'dipolar effect' of the $C_2$  axis
averages out in the $D_{2h}$-symmetric nematic phase and
the symmetry of the
fluctuating V-shaped model in the mean-field approximation
is effectively that of  the fluctuating quadrupole model.
\end{thebibliography}
\end{document}